\documentclass[preprint,12pt]{elsarticle}

\usepackage{amssymb, amsmath}
\usepackage{booktabs, hhline}
\usepackage{rotating}
\usepackage[normalem]{ulem}

\newcommand {\rot}[1]  {\begin{sideways} #1 \end{sideways}}

\journal{Nuclear Physics A}

\begin{document}

\begin{frontmatter}

%% Title, authors and addresses

%% use the tnoteref command within \title for footnotes;
%% use the tnotetext command for theassociated footnote;
%% use the fnref command within \author or \address for footnotes;
%% use the fntext command for theassociated footnote;
%% use the corref command within \author for corresponding author footnotes;
%% use the cortext command for theassociated footnote;
%% use the ead command for the email address,
%% and the form \ead[url] for the home page:
%% \title{Title\tnoteref{label1}}
%% \tnotetext[label1]{}
%% \author{Name\corref{cor1}\fnref{label2}}
%% \ead{email address}
%% \ead[url]{home page}
%% \fntext[label2]{}
%% \cortext[cor1]{}
%% \affiliation{organization={},
%%             addressline={},
%%             city={},
%%             postcode={},
%%             state={},
%%             country={}}
%% \fntext[label3]{}

\title{Constraining the chirally motivated $\pi\Sigma - \bar{K}N$ models 
with the $\pi\Sigma$ photoproduction mass spectra}

%% use optional labels to link authors explicitly to addresses:
%% \author[label1,label2]{}
%% \affiliation[label1]{organization={},
%%             addressline={},
%%             city={},
%%             postcode={},
%%             state={},
%%             country={}}
%%
%% \affiliation[label2]{organization={},
%%             addressline={},
%%             city={},
%%             postcode={},
%%             state={},
%%             country={}}

\author{A.~Ciepl\'{y} \corref{correspondence}}
\author{P.~C.~Bruns}
\cortext[correspondence]{Corresponding author}
\ead{cieply@ujf.cas.cz}
\address{Nuclear Physics Institute of the Czech Academy of Sciences, 250 68 \v{R}e\v{z}, Czechia}

\begin{abstract}
The paper presents a first time attempt on a combined fit of the $K^{-}p$ low-energy data and the $\pi\Sigma$ photoproduction 
mass spectra, performed without fixing the meson-baryon rescattering amplitudes to a specific $\pi\Sigma - \bar{K}N$ coupled channels 
model obtained from fitting exclusively the $K^{-}p$ data. The formalism adopted to describe the photoproduction process 
is based on chiral perturbation theory and employs a limited number of free parameters. The achieved description 
of the photoproduction mass distributions is not quite satisfactory, leaving a room for improving the photo-kernel construction, 
but still provides additional constraints on the positions of the $\Lambda(1405)$ poles. In particular, the presented 
models tend to limit the mass of the lower pole and yield a larger width of the $\bar{K}N$ related pole at a higher mass.
\end{abstract}

\begin{keyword}
%% keywords here, in the form: keyword \sep keyword
meson-baryon interactions \sep chiral perturbation theory \sep $\Lambda(1405)$ \sep $\pi\Sigma$ photoproduction 
%% PACS codes here, in the form: \PACS code \sep code
%\PACS 0000 \sep 1111
%% MSC codes here, in the form: \MSC code \sep code
%% or \MSC[2008] code \sep code (2000 is the default)
%\MSC 0000 \sep 1111
\end{keyword}

\end{frontmatter}

%% \linenumbers

%%%%%%%%%%%%%%%%%%%%%%%%%%%%%%%%%%%%%%%%%%%%%%%%%%%%%%%%%%%%%%%%%%%%%%%%%%%%%%%%%%%%%%%%%%%%%%%
\section{Introduction}
\label{sec:intro}
%%%%%%%%%%%%%%%%%%%%%%%%%%%%%%%%%%%%%%%%%%%%%%%%%%%%%%%%%%%%%%%%%%%%%%%%%%%%%%%%%%%%%%%%%%%%%%%

The modern theoretical approaches to strangeness $S=-1$ meson-baryon ($MB$) scattering, which are based 
on a combination of chiral effective Lagrangians with non-perturbative coupled-channel formalism, 
have been quite successful in describing low-energy $K^{-}p$ data 
\cite{Kaiser:1995eg, Jido:2003cb, Ikeda:2012au, Cieply:2011nq, Guo:2012vv, Mai:2014xna, Feijoo:2018den}. 
However, due to the non-perturbative 
extension of the chiral amplitudes, the uncertainty of the values of higher-order low-energy constants (LECs) 
in the effective Lagrangians, and the lack of data for processes like e.g.~elastic $\pi\Sigma$ scattering, 
there remains a substantial model dependence in the description of the subthreshold antikaon-nucleon amplitude, 
and in particular in the determination of the pole position(s) of the subthreshold $\Lambda(1405)$ 
resonance(s), see \cite{Hyodo:2011ur,Mai:2020ltx} for reviews on this subject. Even though 
it is probably fair to say that most modern analyses agree that there are two poles on the complex-energy 
surface related to the $\Lambda(1405)$ \cite{Oller:2000fj,Cieply:2016jby,Meissner:2020khl}, 
and the position of the narrow resonance just below the $\bar{K}N$ threshold seems to be well fixed 
by the $K^{-}p$ data, the position and nature of the assumed second resonance are still poorly constrained. 
A possible way to reduce this model dependence was provided by the measurement of the $\gamma p\rightarrow K^{+}\pi\Sigma$ 
photoproduction cross sections by the CLAS collaboration. In \cite{CLAS:2013rjt}, the $\pi\Sigma$ line shapes 
in the $\Lambda(1405)$ resonance region were determined, and in \cite{CLAS:2013rxx}, the $K^{+}$ angular 
distributions were measured as well.
%(see also \cite{Ahn:2003mv,Niiyama:2008rt,Lu:2013nza}).
The pole structure in the meson-baryon scattering amplitudes is expected to influence the $\pi\Sigma$ line shapes notably through the final-state interaction (FSI) of the produced mesons and baryons, so that it should be possible to rule out classes of models which are not consistent with the CLAS data. The theoretical question is then how to implement the FSI given by on-shell meson-baryon partial-wave scattering amplitudes in the two-meson photoproduction amplitude. A pragmatic approach was followed in \cite{Roca:2013cca,Mai:2014xna}, where the elementary photoproduction amplitude was parameterized by a large set of constants $C^{c}(\sqrt{s})$ (for each reaction channel $c$, and each bin of the c.m. energy $\sqrt{s}$). Multiplying this by a meson-baryon loop function and a unitarized coupled-channel meson-baryon scattering amplitude to describe the FSI, one obtains a simple model parameterization for the full production amplitude.  Comparing the quality of fits to the CLAS data with this class of models then allows to test the employed scattering amplitudes. The drawbacks of such an approach are, of course, that it is hard to judge whether the obtained fit results for the $C^{c}(\sqrt{s})$ make sense, and that one does not arrive at a more detailed understanding of the full process. To achieve the latter, microscopic models for the elementary photoproduction amplitude, based on effective Lagrangians, have been developed \cite{Nacher:1998mi,Nakamura:2013boa,Bruns:2020lyb} (see also \cite{Lutz:2004sg,Wang:2016dtb,Nam:2017yeg}). In a recent contribution \cite{Bruns:2022sio}, we have started our analysis of the CLAS data by utilizing the formalism presented in \cite{Bruns:2020lyb} and the meson-baryon scattering amplitudes constructed in \cite{Mai:2014xna,Sadasivan:2018jig,Bruns:2021krp}, which were treated as a {\em fixed\,} FSI in the formalism, i.~e. these amplitudes were not refitted to the CLAS data to allow for a test of their respective predictions. 
%The previous fits of the $\pi\Sigma$ mass distributions provided by the CLAS experiment \cite{CLAS:2013rjt} 
%were performed with the meson-baryon rescattering in the final state generated by $\pi\Sigma - \bar{K}N$ 
%coupled channels approaches that had their parameters already fixed in fits to the $K^{-}p$ reactions data. 
%However, the available $\pi\Sigma - \bar{K}N$ models yield quite varied predictions when extended to energies 
%below the $K^{-}p$ threshold (see the comparison made in \cite{Cieply:2016jby}) and it was also 
%demonstrated in \cite{Bruns:2022sio} that this model dependence applies to the analysis of the CLAS data as well.
As it was found in \cite{Bruns:2022sio} that the FSI plays a major role in the description of the $\pi\Sigma$ photoproduction mass spectra, it seems a 
natural next step to use the CLAS data to restrict the parameter space of the $\pi\Sigma - \bar{K}N$ models 
by fitting the latter not only to the $K^{-}p$ reactions data but also to the $\pi\Sigma$ mass distributions. 
This is exactly what we have done, and the results of our analysis represent the main content of this paper.

In the next section, we shortly recall the formalism of \cite{Bruns:2020lyb,Bruns:2022sio} to be used in this contribution, 
and describe some minor modifications to it aimed at a better consistency of the photoproduction model with the amplitudes employed 
for the FSI. In Sec.~\ref{sec:results}, we present the results of our analysis in detail, and in Sec.~\ref{sec:conclusions}, 
we offer the conclusions drawn from those results.

%%%%%%%%%%%%%%%%%%%%%%%%%%%%%%%%%%%%%%%%%%%%%%%%%%%%%%%%%%%%%%%%%%%%%%%%%%%%%%%%%%%%%%%%%%%%%%%
\section{Formalism}
\label{sec:formalism}
%%%%%%%%%%%%%%%%%%%%%%%%%%%%%%%%%%%%%%%%%%%%%%%%%%%%%%%%%%%%%%%%%%%%%%%%%%%%%%%%%%%%%%%%%%%%%%%

\begin{figure}[t]
\centering
\includegraphics[width=0.4\textwidth]{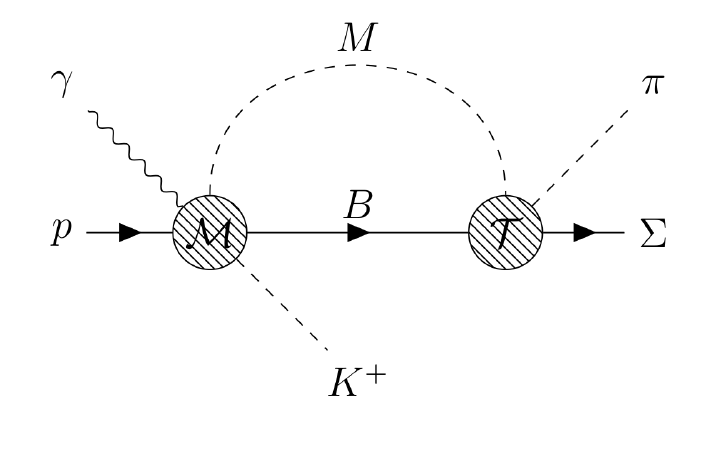}\\
\vspace*{-5mm}
\caption{Illustration of the final-state interaction (FSI) of the meson-baryon pair. $\mathcal{M}$ represents the amplitude 
for the $\gamma p \to K^{+}MB$ without FSI, and $\mathcal{T}$ is the meson-baryon scattering amplitude, which can be decomposed 
into partial waves $f_{\ell\pm}$.}
\label{fig:MBFSI}  
\end{figure}

The $MB$ coupled-channel models mentioned in the previous section provide us with partial-wave 
scattering amplitudes $f_{\ell\pm}^{c',\,c}(M_{\pi\Sigma})$ that describe the transitions  
from channel $c$ to channel $c'$, where the indices represent the charged states of 
$\pi\Lambda,\, \pi\Sigma,\, \bar{K}N,\, \eta\Lambda,\, \eta\Sigma$ and $K\Xi$, 10 channels in total. 
These amplitudes are designed to be consistent with ChPT up to a fixed order of the low-energy 
expansion and fulfill the requirement of coupled-channel unitarity above the lowest reaction threshold.
They are functions of the two-body c.m.~energy (invariant $MB$ mass) $M_{\pi\Sigma}$ and projected on 
a total angular momentum $\ell\pm\frac{1}{2}$ and orbital angular momentum \mbox{$\ell=0,1,2,\ldots\,$}. 
Considering the photoproduction reaction $\gamma p \to K^{+}MB$ as illustrated in Fig.~\ref{fig:MBFSI}, 
the s-wave scattering amplitudes $f_{0+}$ can be implemented to describe the final-state interaction 
of the produced $MB$ pair. It was shown in Refs.~\cite{Bruns:2020lyb, Bruns:2022sio} that one can 
obtain four independent photoproduction amplitudes $\mathcal{A}^{i(\mathrm{tree})}_{0+}(s,M_{\pi\Sigma})$
that combine the structure functions contributing to the tree-level photoproduction amplitude $\mathcal{M}$ 
which is projected on the $\ell=0$ state of the $\pi\Sigma$ pair. Since the outgoing $K^+$ can be treated effectively 
as a spectator particle in the unitarization procedure, the FSI is then restricted to the $S=-1$ meson-baryon 
subspace, the sector relevant for the formation of the $\Lambda(1405)$. We also note that the $\mathcal{A}$ 
amplitudes depend also on $t_K$ (the squared momentum transfer from the photon to the outgoing spectator kaon) 
which we do not show explicitly to keep the notation concise.

The unitarized photoproduction amplitudes for $\gamma p \to K^{+}MB$ are taken as the coupled-channel vector
\begin{equation}
\label{eq:modelA}
\begin{split}    
[\mathcal{A}^{i}_{0+}(s,M_{\pi\Sigma})] &= [\mathcal{A}^{i(\mathrm{tree})}_{0+}(s,M_{\pi\Sigma})] \\
  &+ [f_{0+}(M_{\pi\Sigma})] \left[8\pi M_{\pi\Sigma}G(M_{\pi\Sigma})\right] [\mathcal{A}^{i(\mathrm{tree})}_{0+}(s,M_{\pi\Sigma})]\, ,
\end{split}
\end{equation}
where $G(M_{\pi\Sigma})$ is a diagonal matrix with entries given by suitably regularized $MB$ loop integrals, and 
the square brackets mark the vector or matrix character of the involved quantities in the $MB$ channel-space. 
The $MB=\pi\Sigma$ entries of the vector $\mathcal{A}^{i}_{0+}$ can then be used to obtain the required s-wave 
cross sections (see Eq.~(2.9) in \cite{Bruns:2022sio}). Our tree-level amplitudes $\mathcal{A}^{i(\mathrm{tree})}_{0+}$ 
are constructed from 16 $\mathcal{M}$-structures representing the Weinberg-Tomozawa, Born, and anomalous graphs adhering 
strictly to constraints arising from unitarity, gauge invariance and chiral perturbation theory, 
see Refs.~\cite{Bruns:2020lyb, Bruns:2022sio} for details. 

With the $MB$ amplitudes fixed by a particular $MB$ model the formalism adopted in \cite{Bruns:2022sio} does not contain 
any free parameter but was found lacking in reproducing the experimental $\pi\Sigma$ line-shapes and completely 
inadequate to emulate the energy dependence yielding too large cross sections at higher energies. 
Thus, we have decided to introduce additional form-factors applied to the photoproduction kernel $\mathcal{M}$ that also help 
with regularization of the intermediate $MB$ loop function connecting the photo-kernel with the rescattering amplitude $\mathcal{T}$ 
and allow to vary to some extent the relative contributions of the various $MB$ states produced in the first step (prior to rescattering) 
of the whole photoproduction process. This adds one additional regularization scale $\beta_{c}$ for each channel $c$ 
in a form factor $g^{\gamma}_{c}$ multiplying the photo-kernel for this channel. These form factors are chosen 
of the same Yamaguchi form as those entering our model for the FSI and denoted $g_{c}$. Then the loop function 
entering Eq.~(\ref{eq:modelA}) is taken as the analogue of the loop function in that model (compare Eqs.~(7) and (8) 
in \cite{Bruns:2019fwi}): 
\begin{equation}
G_{c} := -4\pi\int\frac{d^3p}{(2\pi)^3}\,\frac{g_{c}(p)g^{\gamma}_{c}(p)}{q_{c}^2 - p^2 + {\rm i}\epsilon} 
       = g_{c}(q_{c})g^{\gamma}_{c}(q_{c})\left[\frac{(\alpha_{c}+{\rm i}q_{c})(\beta_{c}+{\rm i}q_{c})}{\alpha_{c}+\beta_{c}}\right]\,,
\end{equation}
where $q_{c}$ is the momentum for channel $c=jb$ (meson $j$, baryon $b$) in the c.m.~frame of the $MB$ pair 
(labeled by an $\ast$ in \cite{Bruns:2020lyb, Bruns:2022sio}), 
\begin{equation}
q_{c}=\frac{1}{2M_{\pi\Sigma}}\sqrt{(M_{\pi\Sigma}^2-(m_{b}+M_{j})^2)(M_{\pi\Sigma}^2-(m_{b}-M_{j})^2)}\,,
\end{equation}
and
\begin{equation}
g_{c}(p) := \frac{\alpha_{c}^2}{\alpha_{c}^2+p^2}\,,\qquad g^{\gamma}_{c}(p) := \frac{\beta_{c}^2}{\beta_{c}^2+p^2}\,.
\label{eq:photoFF}
\end{equation}
Replacing $g^{\gamma}_{c}$ by $g_{c}$, i.e. $\beta_{c}$ by $\alpha_{c}$, reduces $G_{c}$ to the loop function used 
in the construction of $\mathcal{T}$ \cite{Bruns:2019fwi, Bruns:2021krp}. In addition, during our analysis we found 
out that we need to introduce one more form factor for the outgoing spectator kaon, $g_{K^{+}}(q_{K})$, which was chosen 
to be of the same functional form, with an adjustable regulator scale $\beta_{K^{+}}$. This form factor influences 
mainly the dependence of the effective photoproduction kernel on the overall c.m.~energy $\sqrt{s}$ and was necessary 
to improve the earlier mentioned energy dependence of the generated cross sections. 

Finally, the formalism of \cite{Bruns:2020lyb} also provides amplitudes $\mathcal{A}^{i}_{1-}$ which couple 
to the $f_{1-}^{c',\,c}(M_{\pi\Sigma})$ $MB$ amplitudes. Unlike in \cite{Bruns:2022sio}, to make the present analysis 
as complete as possible at this stage, we decided to include the $\mathcal{A}^{i}_{1-}$ amplitudes on the tree level, 
which yields a relatively small background contribution. Concerning the p-wave $MB$ form-factors applied 
to the photo-kernel, we follow \cite{Cieply:2015pwa} and adopt 
$g^{\gamma}_{1c}(p):=\left(1+(p^2/\beta_{c}^2)\right)^{-3/2}$ for the $\mathcal{A}^{i}_{1-}$ projection of the kernel.
Since we focus on the s-wave resonance sector and do not foresee a significant $MB$ rescattering in the $1-$ partial 
wave, the FSI is neglected in this p-wave sector. While, in principle, the FSI is quite relevant in the $1+$ partial wave 
due to the existence of the $\Sigma(1385)$ resonance, one should point out that the effects caused by the latter resonance 
have already been subtracted from the mass spectra of \cite{CLAS:2013rjt} to which we fit our models.
For these reasons, we leave the full treatment of the p-wave amplitudes $\mathcal{A}^{i}_{1\pm}$ for a future work.

%%%%%%%%%%%%%%%%%%%%%%%%%%%%%%%%%%%%%%%%%%%%%%%%%%%%%%%%%%%%%%%%%%%%%%%%%%%%%%%%%%%%%%%%%%%%%%%
\section{Results and discussion}
\label{sec:results}
%%%%%%%%%%%%%%%%%%%%%%%%%%%%%%%%%%%%%%%%%%%%%%%%%%%%%%%%%%%%%%%%%%%%%%%%%%%%%%%%%%%%%%%%%%%%%%%

As we declared in the Introduction the purpose of the present work is to provide additional constraints 
on the $\pi\Sigma - \bar{K}N$ coupled channels models by using them in fits of the CLAS data on $\pi\Sigma$ 
mass distributions observed in the photoproduction reactions off proton. While in \cite{Bruns:2022sio} 
several different theoretical approaches were tested providing varied predictions for the FSI, here we adopt
the meson-baryon Prague model of Ref.~\cite{Bruns:2021krp} and re-fit its parameters to the combined set 
of the experimental data composed of the CLAS data (the $\pi^{0}\Sigma^{0}$, $\pi^{-}\Sigma^{+}$ and 
$\pi^{+}\Sigma^{-}$ mass distributions at $\sqrt{s} = 2.1$ GeV \cite{CLAS:2013rjt}) 
and the data on the $K^{-}p$ reactions (kaonic hydrogen characteristics \cite{SIDDHARTA:2011dsy}, 
$K^{-}p$ threshold branching ratios, and low-energy total cross sections to eight $MB$ channels), 
see \cite{Bruns:2021krp} for a complete specification of the latter and references to the experimental papers. 
The free parameters of the Prague model are the LECs represented by the couplings of the NLO meson-baryon chiral 
Lagrangian (four $d$-couplings, $b_0$, and $b_F$) and six inverse ranges $\alpha_{c}$ used to regularize 
the $MB$ loop functions. Some other parameters are kept fixed at their {\it physical values}: 
$b_D = 0.1$ GeV$^{-1}$, meson decay constants $F_\pi = 92.3$ MeV, $F_K = 1.193 \cdot F_\pi = 110.1$ MeV, 
$F_\eta = 1.28 \cdot F_\pi = 118.1$ MeV, and $MB$ axial couplings $F = 0.46$, $D = 0.80$. This setting 
reduces significantly the number of fitted parameters and speeds up the fits. In total, there are 12 parameters 
in the FSI sector represented by the $\mathcal{T}$ amplitude in Fig.~\ref{fig:MBFSI} that are to be fitted to the data.
These are complemented by the $\beta_{c}$ and $\beta_{K^{+}}$ parameters introduced in the previous section 
to adjust the photoproduction kernel $\mathcal{M}$ via the form-factors given by Eq.~(\ref{eq:photoFF}). 
Since the physical observables at energies around the $\Lambda(1405)$ region should not be affected 
by the virtual $MB$ channels opening high above the $\bar{K}N$ threshold, we fix their respective $\beta$-scales to 
$\beta_{\eta \Lambda} = \beta_{\eta \Sigma} = \beta_{K \Xi}$ = 650 MeV, a {\it natural value} for all four  
channels and close to the rough estimates given by Eq.~(2.13) in Ref.~\cite{Bruns:2022sio}. We also assume that 
the $\beta$ inverse ranges used in the p-wave form-factors can be set equal to their s-wave counterparts. 
After completing our analysis, we checked that allowing for reasonable variations of these pre-fixed scales 
does not have a notable impact on our results. This leaves us with three $\beta_{c}$ and the $\beta_{K^{+}}$ 
scales to be fitted to the data bringing the total number of free parameters to 16. 

The Table~\ref{tab:data} shows the observables we fit and the pertinent numbers of experimental data. 
While the photoproduction mass distributions and the $K^{-}p$ cross sections into different final state 
channels are treated separately, we take the three branching ratios at the $K^{-}p$ threshold 
as one set of data. One should also note that we include the new and very accurate data on the $\pi^0\Lambda$ 
and $\pi^0\Sigma^0$ total cross sections measured by the AMADEUS collaboration \cite{Piscicchia:2022wmd},
the first data below 100 MeV/c kaon momentum that were not used in previous fits of the $\pi\Sigma - \bar{K}N$ 
coupled channels models. The experiment has added one new data point for each of the two final 
states that are relevant for separating the isovector and isoscalar parts of the $MB$ interactions. 
We also mention that we decided to include in our fits only the CLAS data available for the $\sqrt{s} = 2.1$ GeV 
c.~m.~energy as we want to see if the model can cope with the energy dependence when applied to reproduce 
the data measured at other energies. Each data set listed in Table~\ref{tab:data} contributes to the total 
$\chi^{2}/{\rm dof}$ with the same weight,
\begin{equation}  
\chi^{2}{\rm /dof} = \frac{\sum_{i}N_{i}}{N_{obs}(\sum_{i}N_{i}-N_{par})} \sum_{i}\frac{\chi^{2}_{i}}{N_{i}} \, ,
\label{eq:chi2}
\end{equation}
where $N_{par}$ denotes the number of fitted parameters, $N_{obs} = 13$ is a number of observables, $N_{i}$ 
is the number of data points for an $i$-th observable, and $\chi^{2}_{i}$ stands for the total $\chi^{2}$ 
computed for the observable. We use the MINUIT routine from the CERN library of FORTRAN codes to minimize 
the $\chi^{2}$ per degree of freedom defined by Eq.~(\ref{eq:chi2}). 

\begin{table}[htb]
\caption{Observables and numbers of pertinent experimental data included in our analysis.
The first three columns are for the $\pi\Sigma$ photoproduction mass distributions, followed by the kaonic hydrogen 
characteristics and the branching ratios, and the next 8 columns are for the $K^{-}p \rightarrow MB$ reactions.}
\begin{center}
\begingroup
\setlength{\tabcolsep}{6pt}
\begin{tabular}{ccc|cc|cccccccc|c}
\multicolumn{3}{c|}{CLAS data} & \multicolumn{2}{c|}{$K^-p$ threshold} & \multicolumn{8}{c|}{$K^-p$ cross sections} &  all \\
\rot{$\pi^0\Sigma^0$} & \rot{$\pi^-\Sigma^+$} & \rot{$\pi^+\Sigma^-$} & \makebox[10mm][c]{\rot{atom}} & \makebox[10mm][c]{\rot{BRs}} & 
\rot{$\pi^0\Lambda$} & \rot{$\pi^0\Sigma^0$} & \rot{$\pi^-\Sigma^+$} & \rot{$\pi^+\Sigma^-$} &
\rot{$K^-p$}         & \rot{$\bar{K}^0 n$}   & \rot{$\eta\Lambda$}   & \rot{$\eta\Sigma^0$}  & \rot{total} \\ \midrule
34 & 30 & 32 &  2 &  3 &
 4 &  4 & 31 & 32 & 
27 & 22 & 24 &  7 & 252  
\end{tabular}    
\endgroup
\end{center}
\label{tab:data}
\end{table}

The results of our fits are shown in Table~\ref{tab:parameters} presenting the values of the fitted parameters 
and the achieved $\chi^{2}{\rm /dof}$ for four selected local minima that represent models tagged as P0, P1, P2 and P3.
In the first P0 model the meson-baryon amplitudes used in the FSI were kept fixed, generated by the Prague model 
specified in \cite{Bruns:2021krp}, and only the four $\beta$ scales were varied. Thus, the pertinent 
$\chi^{2}{\rm /dof} = 5.40$ is higher and not directly comparable with the values reached in the other fits. 
Having the LECs and $\alpha$ scales fixed also means that the P0 model generates exactly the same results 
in the sector of $K^{-}p$ reactions as the original Prague model of Ref.~\cite{Bruns:2021krp}. The other three 
models discussed in the current work were obtained while varying all 16 parameters, the P1 one representing
a global $\chi^{2}$ minimum and P2 a local minimum that we got while using the P0 model parameters as a starting 
point for running the fitting procedure.

\begin{table}[htb]
\caption{The inverse regularisation ranges (in MeV) and NLO couplings (in GeV$^{-1}$) obtained 
in fits of the combined $\pi\Sigma$ mass distributions and $K^{-}p$ data. The values marked with 
an asterisk were taken from Ref.~\cite{Bruns:2021krp} and kept fixed in the P0 fit. In addition,
we have also preset the parameters  $b_D=0.100$ GeV$^{-1}$ and 
$\beta_{\eta \Lambda} = \beta_{\eta \Sigma} = \beta_{K \Xi}$ = 650 MeV. 
The last line provides the values of the $\chi^{2}/{\rm dof}$.}
\begin{center}
\begin{tabular}{c|rrrr}
 parameter               &  P0          &  P1    &  P2    &  P3    \\ \noalign{\smallskip} \hline \hline \noalign{\smallskip}
 $b_0$                   &  0.525$^{*}$ & -0.833 & -0.596 & -0.387 \\
 $b_F$                   & -0.077$^{*}$ & -0.061 & -0.106 & -0.036 \\
 $d_1$                   & -0.119$^{*}$ &  0.835 & -0.169 & -0.084 \\
 $d_2$                   &  0.074$^{*}$ & -0.275 &  0.037 &  0.001 \\
 $d_3$                   &  0.096$^{*}$ & -0.637 &  0.026 & -0.155 \\
 $d_4$                   &  0.556$^{*}$ &  0.481 & -0.529 & -0.423 \\
 $\alpha_{\pi \Lambda}$  &    400$^{*}$ &    322 &    276 &    301 \\
 $\alpha_{\pi \Sigma}$   &    509$^{*}$ &   1118 &    480 &    647 \\
 $\alpha_{\bar{K}N}$     &    752$^{*}$ &    467 &    809 &   1109 \\
 $\alpha_{\eta \Lambda}$ &    979$^{*}$ &   1217 &    500 &   1500 \\
 $\alpha_{\eta \Sigma}$  &    797$^{*}$ &    500 &    500 &    558 \\
 $\alpha_{K \Xi}$        &   1079$^{*}$ &    778 &    815 &   1063 \\ \midrule
 $\beta_{\pi \Lambda}$   &    558       &    455 &   1500 &   1500 \\
 $\beta_{\pi \Sigma}$    &    205       &    207 &    212 &    217 \\
 $\beta_{\bar{K}N}$      &   1393       &    606 &   1500 &    931 \\
 $\beta_{K^+}$           &    930       &   1250 &    696 &    697 \\ \midrule 
 $\chi^{2}/{\rm dof}$    &    5.40      &   3.34 &   4.41 &   4.72 \\ \midrule
\end{tabular}    
\end{center}
\label{tab:parameters}
\end{table}

The form-factors $g_{c}(p)$ play a role of channel filters either effectively enhancing or reducing the contribution 
that comes from photoproduction of the intermediate $MB$ pair, the channel $c$ from which the rescattering FSI kicks off. 
For large values of $\beta_c$ the form factor remains close to 1 while for small values of $\beta_c$ the contribution 
of the pertinent channel is suppressed above its threshold (and enhanced below it). We note that form-factors 
of the same design were introduced in \cite{Nakamura:2013boa}, though there the authors opted for a uniform 
regulator scale adjusted to the data, $\beta_{c} = \beta_{K^{+}} = 511$ MeV. In our scheme, almost all the local 
minima we found with $\chi^{2}/{\rm dof} < 5$ retain $\beta_{K^{+}} \approx 700 - 800$ MeV and profess 
$\beta_{\pi\Sigma}$ values close to 200 MeV. The $\beta_{\bar{K}N}$ values appear much larger, often reaching 
the 1500 MeV limit we have preset as a maximum for the parameter. This means that a $\pi\Sigma$ photoproduction 
is suppressed in the intermediate state, effectively enhancing the role of the higher (in mass) $\Lambda(1405)$ 
pole that couples more strongly to the $\bar{K}N$ channel. 

In Table~\ref{tab:poles} we show the positions of the poles assigned to the dynamically generated resonances 
$\Lambda(1405)$ and $\Lambda(1670)$. All poles are located at Riemann sheets (RS) that are reached by crossing 
the real axis at the resonance mass when going from the physical RS on the complex energy manifold. 
Adopting the notation used e.g.~in Refs.~\cite{Cieply:2016jby, Bruns:2021krp}, the poles $z_1$ and $z_2$ 
are located on the $[-,+,+,+]$ RS and the $z_3$ pole on the $[-,-,-,+]$ RS with the signs referring to 
either physical ($+$) or unphysical ($-$) signs of the imaginary parts of the meson momenta (in the two body 
$MB$ c.m. system) for the isoscalar channels $\pi\Sigma$, $\bar{K}N$, $\eta\Lambda$ and $K\Xi$ (in this order 
of channel thresholds). It is anticipated that the approaches based on the chiral meson-baryon Lagrangian generate 
two poles for the $\Lambda(1405)$ resonance \cite{Oller:2000fj,Cieply:2016jby,Meissner:2020khl}, 
a feature in accordance with the approximate SU(3) flavor symmetry \cite{Bruns:2021krp}. 
Thus, it came to us as a surprise that the P1 model does not follow the suit and 
does not generate the $z_1$ and $z_3$ poles where one would expect them. To find out why the poles disappear 
we followed the movement of the $\Lambda(1405)$ poles while gradually modifying the parameter setting from 
the P0 to the P1 one. The result is that while the $z_2$ pole smoothly changes its position from the P0 
to the P1 one, the $z_1$ pole moves to lower energies, reaches the real axis and then continues moving along it 
to end up at energy well below 1200 MeV, very far from the physical region. In addition, a more closer examination 
of the P1 solution also revealed that it generates an isovector pole at the energy $z_4=(1324, -1.84)$, 
quite close to the real axis on the $[-,+,+,+,+]$ RS (note, that there are 5 channels in the $I=1$ sector). 
This pole leads to a narrow spike in the $MB$ amplitudes involving the isovector component which we find 
unacceptable, as no such extremely narrow isovector s-wave state is known in the energy region in question.
Therefore, we deem the P1 solution as unphysical, though we keep it in the discussion of our results to demonstrate 
the pitfalls one faces when fitting the model parameters to the experimental data. 

\begin{table}[htb]
\caption{The positions of the poles assigned to the dynamically generated $\Lambda(1405)$ and $\Lambda(1670)$ 
resonances. The P1 model generates only one isoscalar pole in the physically relevant region.}
\begin{center}
\begin{tabular}{c|ccc}
 model & $z_1$ [MeV] & $z_2$ [MeV] & $z_3$ [MeV] \\ \midrule
 P0    & (1353,-43)  & (1428,-24)  & (1677,-14)  \\
 P1    &    ---      & (1421,-43)  &    ---      \\
 P2    & (1347,-71)  & (1425,-46)  & (1725,-57)  \\ 
 P3    & (1345,-58)  & (1425,-45)  & (1665,-7.1) \\ \midrule
\end{tabular}    
\end{center}
\label{tab:poles}
\end{table}

\begin{figure}[htb]
  \centering
  \includegraphics[width=0.5\textwidth]{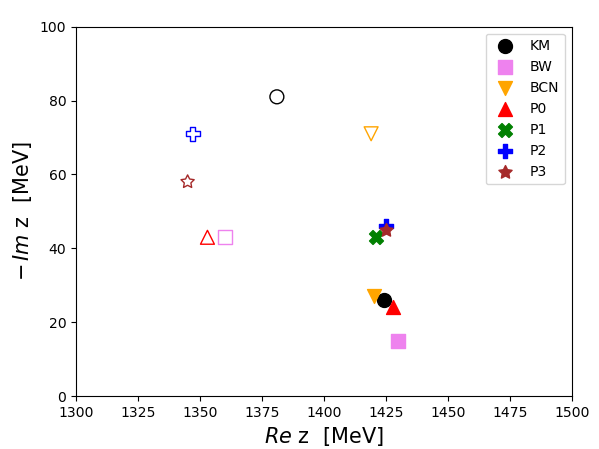} 
\caption{The $\Lambda(1405)$ poles generated dynamically within the meson-baryon coupled channels models. 
The predictions of the Kyoto-M\"unchen (KM) \cite{Ikeda:2012au}, Bonn-Washington (BW) \cite{Sadasivan:2018jig} 
and Barcelona (BCN) \cite{Feijoo:2018den} groups are shown for a comparison too.
}
\label{fig:poles}
\end{figure}

The Figure~\ref{fig:poles} shows the $z_1$ and $z_2$ pole positions in comparison with predictions 
of other $\pi\Sigma - \bar{K}N$ coupled channels approaches. There, the $z_2$ poles that couple 
more strongly to the $\bar{K}N$ channel and (usually) have a higher {\it mass} are marked with filled-in 
symbols while their $z_1$ cousins are marked with the empty symbols of the same form. It is interesting 
to note the clustering of the P1, P2 and P3 $z_2$ pole positions at about the same complex energy but 
with about twice as large imaginary part when compared with the P0 model prediction (and those by other groups) 
that was based solely on fits of the $K^{-}p$ reactions data. The predictions of the $z_1$ position 
also appear to be more constrained than in previous analyses aiming exclusively at reproducing 
the $K^{-}p$ data. Although we show here only the P2 and P3 positions, we have checked 
that several other local minima with $\chi^{2}/{\rm dof} < 5$ also generate the $z_1$ resonance
at a mass around 1350 MeV and width (with a slightly more varied value) of about 120 MeV. 

\begin{figure}[htb]
  \begin{minipage}[t]{0.5\textwidth}
    \includegraphics[width=\textwidth]{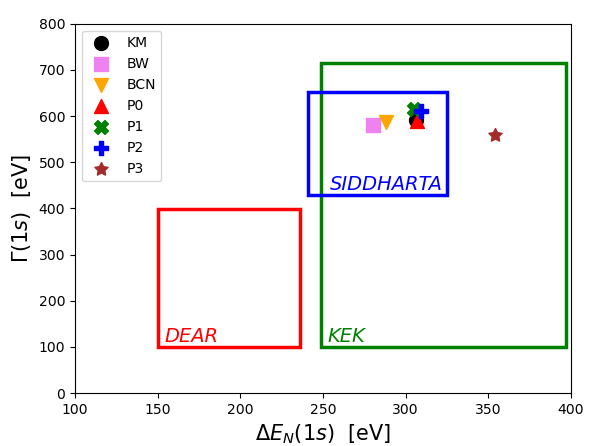} 
  \end{minipage}
  \begin{minipage}[b]{0.49\textwidth}
     \includegraphics[width=\textwidth]{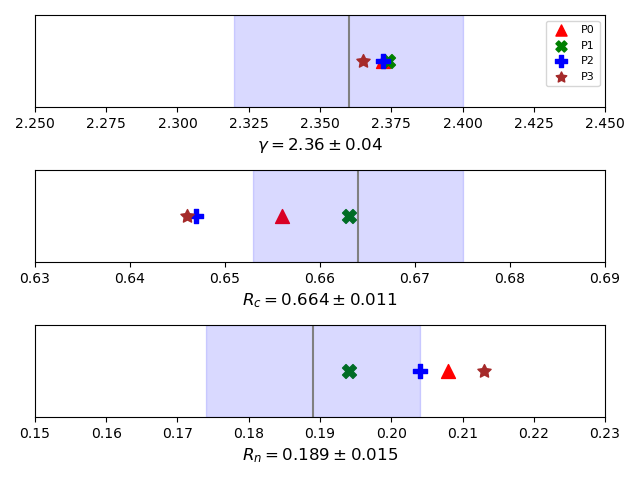}
  \end{minipage}
\caption{The reproduction of the $K^{-}p$ threshold data. Left panel: The theoretical predictions of the kaonic hydrogen 
1s level energy shift $\Delta E_N(1s)$ and width $\Gamma (1s)$ caused by strong interaction are shown for the same models 
as in Fig.~\ref{fig:poles}. The rectangle areas cover the regions within one standard deviation 
for the KEK \cite{Ito:1998yi}, DEAR \cite{DEAR:2005fdl} and SIDDHARTA \cite{SIDDHARTA:2011dsy} experiments.
Right panel: The reproduction of the three threshold branching ratios specified in Eq.~(\ref{eq:BRs}). 
Here, the shaded areas display the intervals withing one standard deviation off the experimental values.}
\label{fig:KNthr}
\end{figure}

The reproduction of the experimental data at the $K^{-}p$ threshold is visualized in Fig.~\ref{fig:KNthr}. 
In the left panel, we compare our results for the kaonic hydrogen characteristics (the 1s level energy shift 
and absorption width due to strong interaction) with those from several previous analyses 
\cite{Ikeda:2012au, Sadasivan:2018jig, Feijoo:2018den}. On the right, we demonstrate how our models 
reproduce the three threshold branching ratios, 
\begin{align}
\gamma & = \frac{\Gamma(K^-p\rightarrow \pi^+\Sigma^-)}{\Gamma(K^-p\rightarrow \pi^-\Sigma^+)} \;,  \nonumber \\
 R_n & = \frac{\Gamma(K^-p\rightarrow \pi^0\Lambda)}{\Gamma(K^-p\rightarrow \text{neutral states})} \;, \label{eq:BRs} \\ 
 R_c & = \frac{\Gamma(K^-p\rightarrow \pi^+\Sigma^-, \pi^-\Sigma^+)}{\Gamma(K^-p\rightarrow \text{inelastic channels})}\;.  \nonumber
\end{align}
There, the predictions of the other groups are not included to avoid overcrowding the plots. In general, 
our description of the $K^{-}p$ threshold data is good with only the P3 model predictions 
being several times more than one standard deviation off the experimental values. It is worth mentioning 
how exceptionally well the unphysical P1 model does when reproducing these rather precise data. 

\begin{figure}[htb]
  \centering
  \includegraphics[width=\textwidth]{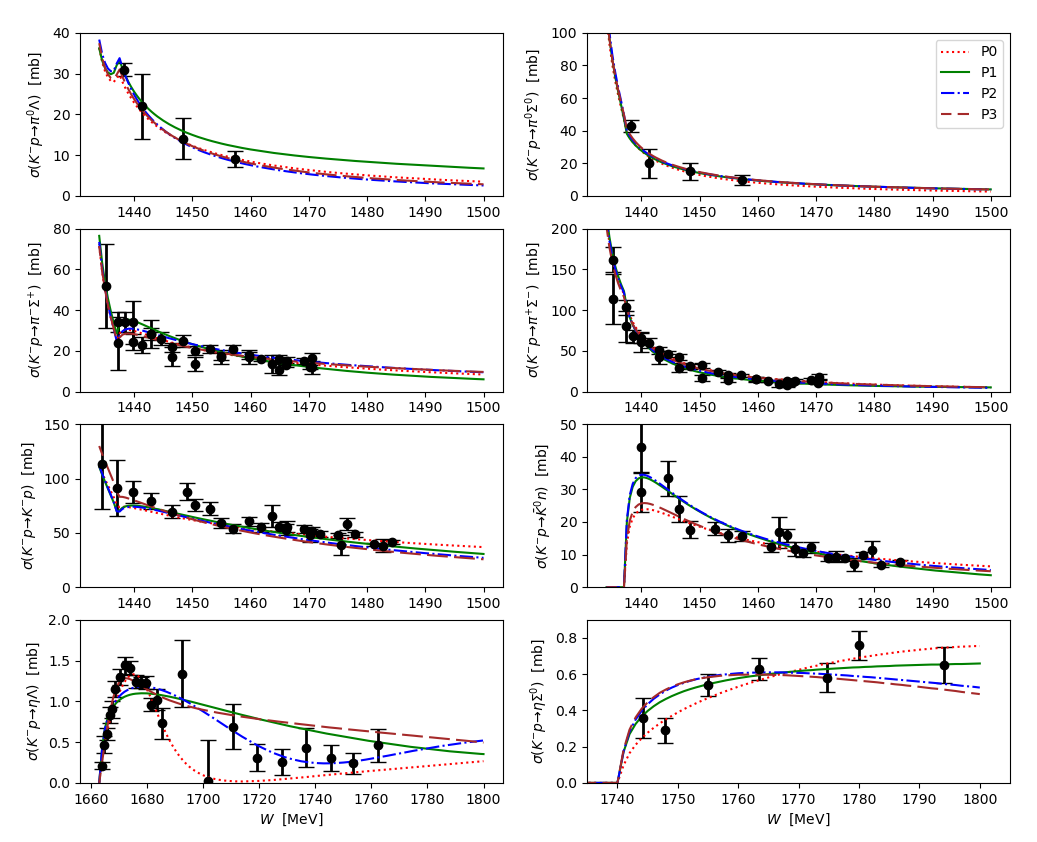} 
\caption{The $K^{-}p$ total cross sections.}
\label{fig:xsections}
\end{figure}

Similarly, the Figure \ref{fig:xsections} demonstrates the quality of reproducing the low-energy total cross sections 
for the $K^{-}p$ reactions. As the experimental data mostly come from relatively old bubble chamber experiments with 
not too restrictive error bars the theoretical models have usually no problem to describe them. The only exception 
is the new AMADEUS measurement \cite{Piscicchia:2022wmd} that added a single data point at the lowest energy 
for the $\pi^0\Lambda$ and $\pi^0\Sigma^0$ cross sections. These new data are really important not only for their 
higher precision but because they come in pure $I=1$ and $I=0$ channels enabling a better separation of isospin 
contributions in the $MB$ coupled channels models. While working on our analysis we have found out that inclusion 
of these new data has helped to eliminate several $\chi^{2}$ local minima of comparable quality to those 
we decided to keep. At the same time the impact of the new data on the presented models was minimal, worsening 
the total $\chi^{2}/{\rm dof}$ by just about $0.1$ and not altering much the positions of the $\Lambda(1405)$ 
poles or any other results discussed here. 

As one can see in Fig.~\ref{fig:xsections} all considered models describe the $K^-p$ cross sections about 
equally well, the only exception being the $\eta\Lambda$ reaction channel. While in Ref.~\cite{Bruns:2021krp} 
the inclusion of these data helped us to get the $\Lambda(1670)$ pole under a better control, the addition 
of the CLAS data puts more weight on the $\Lambda(1405)$ energy region and the position of the $\Lambda(1670)$ 
pole becomes less constrained. This is reflected in Table \ref{tab:poles}, where the $z_3$ pole energies 
(besides the P0 model of \cite{Bruns:2021krp}) do not relate well to the PDG values of the $\Lambda(1670)$ resonance: 
mass $m \approx 1675$ MeV, full width $\Gamma \approx 30$ MeV. It is also nicely seen in Fig.~\ref{fig:xsections}, 
where the $K^-p \rightarrow \eta\Lambda$ peak at lowest energies is reproduced perfectly only by the P0 model 
and  to some extent by the P3 model. The P1 model does not generate the $z_3$ pole at all (and no resonance 
structure is formed) while the P2 model has the pole shifted to higher energies and professing a much larger 
width, both features reflected by the pertinent lineshape in Fig.~\ref{fig:xsections}. Leaving aside the 
$\eta\Lambda$ channel that opens at quite higher energy we find it disturbing that the low-energy $K^{-}p$ 
data (including the threshold ones) cannot clearly distinguish among models professing one or two poles 
assigned to $\Lambda(1405)$ as the unphysical one-pole P1 model describes these data about equally well 
as the other two-poles models do.

\begin{figure}[htb]
  \centering
  \includegraphics[width=\textwidth]{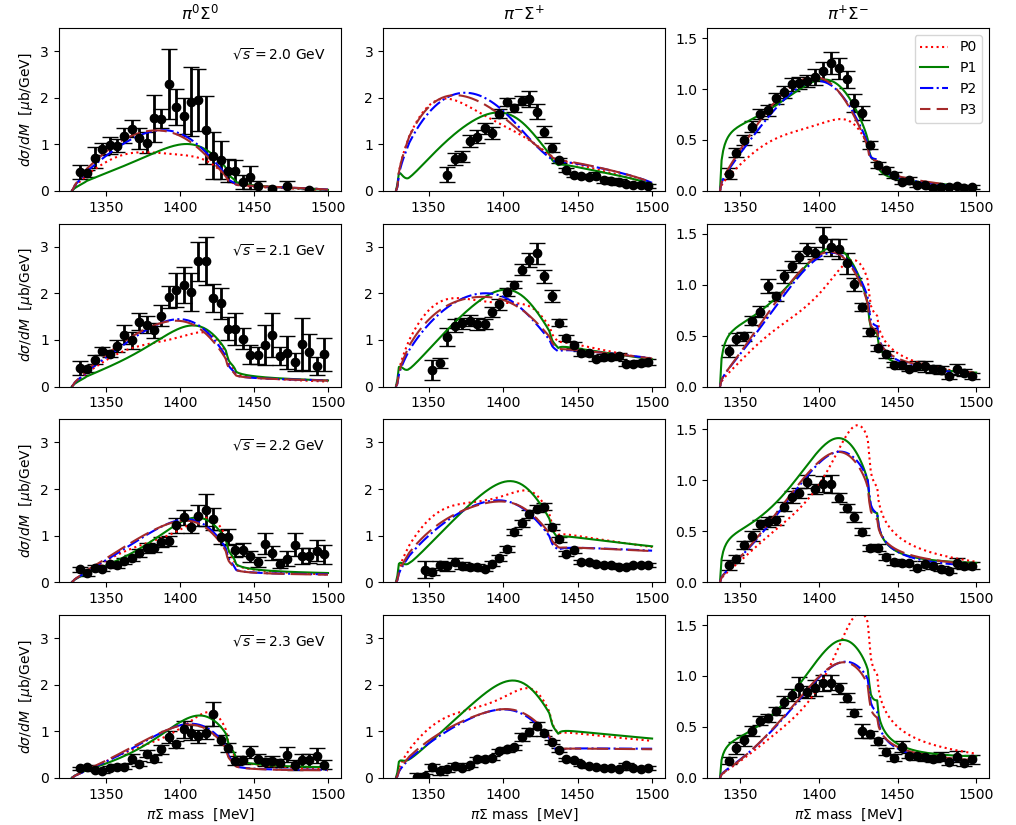} 
\caption{A comparison of the generated $\pi\Sigma$ mass distribution with the CLAS data. The subfigures are arranged 
in rows and columns according to the pertinent c.m.~energy $\sqrt{s}$ and channel, respectively.}
\label{fig:CLAS}
\end{figure}

Finally, we move our attention to the $\pi\Sigma$ mass spectra in the photoproduction reaction on proton. 
In Fig.~\ref{fig:CLAS} we show how the considered four models reproduce the CLAS experimental data \cite{CLAS:2013rjt}
for the c.m.~energies $\sqrt{s} \leq 2.3$ GeV. We remind the reader that only the experimental data at  $\sqrt{s} = 2.1$ GeV 
were used in our fits while the theoretical predictions are shown for all energies reported by the CLAS collaboration 
with the exception of the highest energy $\sqrt{s} = 2.4$ GeV for which the emitted $K^{+}$ momenta 
($|\vec{q}_{K}| \approx 717$ MeV) are considered too high for an approach based on effective field theory. 
Considering that our treatment of the photo-kernel, the $\mathcal{M}$ amplitude in Fig.~\ref{fig:MBFSI}, is relatively 
simple and only four parameters (the $\beta$ scales) were fitted besides those already used in the $\pi\Sigma-\bar{K}N$ 
coupled channels model, the description of the $\pi^{0}\Sigma^{0}$ and $\pi^{+}\Sigma^{-}$ mass distributions 
is quite reasonable. When compared with the results presented in \cite{Bruns:2022sio} we managed to get 
under control the energy dependence, i.e.~the magnitude of the generated spectra. To a large extent, this success 
is due to introducing the form-factor $g^{\gamma}_{K^+}(p)$ related to the kaon emission. Our attempts to work 
without this energy moderator were not successful leading to too large cross sections at energies $\sqrt{s} > 2.1$ GeV
and smaller cross sections at $\sqrt{s} = 2.0$ GeV. 

We admit that our model fails to reproduce the $\pi^{-}\Sigma^{+}$ mass distributions. This is quite in contrast 
with the relatively good reproduction of the $\pi^{+}\Sigma^{-}$ spectra and indicates that the adopted model misses 
some contributions that would combine their isovector and isoscalar parts in a way to satisfy simultaneously 
the experimental data for both charged $\pi\Sigma$ channels. We also have no explanation why the $\pi^{-}\Sigma^{+}$ 
cross sections reported by the CLAS collaboration seem to start about 30 MeV above the threshold while our models 
predicts a steep rise right from the threshold energy. This steep increase is a common feature of all models 
discussed here, not only of the P1 one that generates an unphysical isovector resonance with a mass close 
to the $\pi\Sigma$ threshold. The impact of this resonance on the $\pi^{-}\Sigma^{+}$ mass distribution just 
causes a bump which can be noted in Fig.~\ref{fig:CLAS} slightly above the reaction threshold. We also recall 
that our theoretical treatment of the photoproduction process revealed one characteristics in which the 
$\pi^{-}\Sigma^{+}$ is special. As it was reported in \cite{Bruns:2022sio} it is the only $\pi\Sigma$ channel 
that has non-negligible mass distributions already at the tree level, without employing the FSI. We have checked 
that this is a result of positive interference between large contributions from several tree level graphs 
while in the other two $\pi\Sigma$ channels the same contributions practically cancel each other 
due to destructive interference. However, it is unclear whether this feature could explain the difficulty 
of reproducing the experimental $\pi^{-}\Sigma^{+}$ mass distributions that are dominated by the FSI.

We close this section with emphasizing the relative simplicity of the adopted formalism presented here. 
The number of parameters fitted to the data is quite low and none of them is energy dependent. In is in contrast 
with previous approaches to the $\gamma p \rightarrow K^{+}\pi\Sigma$ photoproduction that introduced 
more graphs in the photo-kernel, including e.g.~exchanges of vector mesons \cite{Nakamura:2013boa} 
or triangle diagrams \cite{Wang:2016dtb}, and still had to fit much larger numbers of free parameters 
(many of them energy dependent) to achieve a good reproduction of the CLAS data. It is thus understandable 
that our current model does not describe the CLAS data so well but our intention for the future is 
to improve the photo-kernel by adding more contributions to it while keeping the number 
of free parameters restricted and under control.

%%%%%%%%%%%%%%%%%%%%%%%%%%%%%%%%%%%%%%%%%%%%%%%%%%%%%%%%%%%%%%%%%%%%%%%%%%%%%%%%%%%%%%%%%%%%%%%
\section{Conclusions}
\label{sec:conclusions}
%%%%%%%%%%%%%%%%%%%%%%%%%%%%%%%%%%%%%%%%%%%%%%%%%%%%%%%%%%%%%%%%%%%%%%%%%%%%%%%%%%%%%%%%%%%%%%%

The main goal of the present work was to study whether the $\pi\Sigma - \bar{K}N$ coupled channels models 
and positions of the poles generated by them can be constrained by performing simultaneous fits of the $K^-p$ data 
together with the $\pi\Sigma$ photoproduction data. Although we have not managed to bring the calculated $\pi\Sigma$ 
mass distributions into satisfactory agreement with the CLAS data, especially in the $\pi^-\Sigma^+$ channel, 
we believe our results already demonstrate the viability of the chosen approach and can provide a guidance 
for future efforts in this direction. When compared with previous studies of the $\pi\Sigma - \bar{K}N$ 
coupled channels system, that concentrated on the reproduction of experimental data in the $K^{-}p$ reactions, 
including the CLAS photoproduction data in fits of the model parameters seems to lead to larger imaginary part 
of the $\Lambda(1405)$ pole that couples more strongly to the $\bar{K}N$ channel, a feature that may relate 
well to an unexpectedly large absorption width found for the $\bar{K}NN$ bound state \cite{J-PARCE15:2020gbh}. 
It also appears that the $\pi\Sigma$ mass distributions put additional constraints on the position of 
{\it the lower mass} $\Lambda(1405)$ pole that is more relevant for the $\pi\Sigma$ photoproduction process. 
The local $\chi^2$ minima found in our work agree on the position of the {\it higher mass pole} at 
$z_2 \approx (1425, -45)$ MeV and the mass of the {\it lower pole} seems to be restricted to the interval 
Re $z_1 \approx 1350 \pm 10$ MeV. The best fit solution P1 yields only one pole in the $\Lambda(1405)$ region 
but was found lacking in several aspects and deemed unphysical, in particular due to generating a narrow 
isovector resonance close to the $\pi\Sigma$ thresholds. However, the fact that the P1 model reproduces 
the $K^{-}p$ data about equally well as the other solutions should make us wary when drawing conclusions 
when interpreting fits to experimental data that do not restrict sufficiently the theoretical models 
used in the analysis. In this respect, we clearly need more new experimental data for the $\bar{K}N$ 
reactions at as low kaon momenta as possible. We hope that the new data reported recently by the AMADEUS 
collaboration \cite{Piscicchia:2022wmd} are just the beginning and more precise low-energy $K^{-}p$ data 
will come in the near future. We are also aware of the new experimental data on the $\pi^0\Sigma^0$ mass distribution 
that were measured by the GlueX collaboration, the preliminary results were already presented in \cite{Wickramaarachchi:2022mhi}. 
Unfortunately, these data come at $\sqrt{s} > 3.6$ GeV, an energy apparently too high for the ChPT treatment adopted 
in our approach.

Since we were not able to describe sufficiently well the $\pi\Sigma$ mass spectra we intend to enhance 
the photoproduction formalism by implementing contributions involving vector mesons (in particular $K^{*}$) 
as well as decuplet baryons and additional resonances in the intermediate state. In this view the current 
work can be regarded as a progress report that demonstrates the capacity of the chosen approach, but 
the presented results should be taken with caution as a more clear picture will emerge with 
the more advanced photo-kernel model.

%% If you have bibdatabase file and want bibtex to generate the
%% bibitems, please use
%%
 \bibliographystyle{elsarticle-num} 
 \bibliography{2023constrainingKN}

\end{document}